\documentclass[useAMS,usenatbib]{pasa}

\input{epsfig.sty}

\def\kms{\ifmmode {\,\rm km \, s^{-1}}
\else {$\rm km \, s^{-1}$}\fi}
\def\Mpc{\ifmmode {\, h^{-1} \, {\rm Mpc}}
\else {$h^{-1}\,$ Mpc}\fi}

\def\s8{{\sigma_8}}

\def\ltsima{$\; \buildrel < \over \sim \;$}
\def\simlt{\lower.5ex\hbox{\ltsima}} 
\def\gtsima{$\; \buildrel > \over \sim \;$} 
\def\simgt{\lower.5ex\hbox{\gtsima}}

\def\omegam{{\Omega_{\rm m}}}

\def\omegab{{\Omega_{\rm b}}}

\def\omegal{{\Omega_\Lambda}}
 
\def\omegabh2{{\omegab h^2}}

\def\s8m{{\sigma_{8{\rm m}}}}
\def\s8g{{\sigma_{8{\rm g}}}}

\def\hompc{\ifmmode {\,h\,\rm Mpc^{-1}}
\else {$h^{-1}$~Mpc}\fi}
\def\m@th{\mathsurround=0pt }
\def\eqalign#1{\null\,\vcenter{\openup1\jot \m@th
 \ialign{\strut\hfil$\displaystyle{##}$&$\displaystyle{{}##}$\hfil
 \crcr#1\crcr}}\,}

\title[2dFGRS as a Cosmological Laboratory]{The 2dF Galaxy Redshift Survey \\
as a Cosmological Laboratory}
\author[Lahav]{Ofer Lahav$^{1,2}$}
\affil{
$^1$Department of Physics and Astronomy , University College London, 
Gower Street, London WC1E 6BT, UK\\
$^2$lahav@star.ucl.ac.uk}
\begin{document}

\maketitle

\label{firstpage}

\begin{abstract}

The 2dF Galaxy Redshift Survey (2dFGRS) of 230,000 redshifts of nearby
($z\sim0.1$) galaxies is now complete.  It has allowed the 2dFGRS team
and others to estimate fundamental cosmological parameters and to
study galaxy intrinsic properties.  Here we highlight three recent key
results from the survey: (i) an upper limit of about 2eV on the total
mass of the three neutrino flavours, and an intriguing reasonable
fitting of the 2dFGRS power spectrum to a Mixed Dark
Matter model without a Cosmological Constant, but with a low Hubble
constant; (ii) the bimodality of the galaxy population in both
spectral parameterisation and in colour; and (iii) the clustering of
different galaxy types and evidence for relative stochastic biasing.

\end{abstract}


\section{Introduction}

Multifibre technology now allows us to measure  redshifts
of millions of galaxies. 
The Anglo-Australian 2
degree Field Galaxy Redshift Survey\footnote{The 2dFGRS Team comprises:
      I.J. Baldry, C.M. Baugh, J. Bland-Hawthorn, T.J. Bridges, R.D. Cannon, 
      S. Cole,
      C.A. Collins,  
      M. Colless,
      W.J. Couch, N.G.J. Cross, G.B. Dalton, R. DePropris, S.P. Driver,
      G. Efstathiou, R.S. Ellis, C.S. Frenk, K. Glazebrook, E. Hawkins, 
      C.A. Jackson,
      O. Lahav, I.J. Lewis, S.L. Lumsden, S. Maddox, 
      D.S. Madgwick, S. Moody, P. Norberg, J.A. Peacock, B.A. Peterson,
      W. Sutherland, K. Taylor. 
For more details on the survey and resulting publications see http://www.mso.anu.edu.au/2dFGRS/}
(2dFGRS)  
measured redshifts for 230,000 galaxies
selected from the APM catalogue. The survey is now complete and publically available.
The median redshift of the
2dFGRS is ${\bar z} \sim 0.1$,  
down to an
extinction corrected magnitude limit of $b_J<19.45$ (Colless et al. 2001). 
A sample of this size allows large-scale structure statistics
to be measured with very small random errors. 
In this review we summarize some recent results 
from the 2dFGRS on clustering and galaxy biasing.
Comprehensive recent reviews are given by Colless (2003) and Peacock (2003).

\section{The Power spectrum of 2dF Galaxies}

  An initial estimate of the convolved, redshift-space power spectrum of the
 2dFGRS has  been determined (Percival et al. 2001)
 for a sample of 160,000 redshifts. 
 On scales $0.02<k<0.15 \hompc$, the data are
 robust and the shape of the power spectrum is not affected by
 redshift-space or non-linear effects, though the amplitude
 is increased by redshift-space distortions.
 Percival et al. (2001), Efstathiou
 et al. (2002) and Lahav et al. (2002)  compared the  
 2dFGRS and CMB
 power spectra, and concluded that they are consistent with each other.

A key assumption in deriving cosmological parameters from redshift surveys is that 
the biasing parameter, 
defined as the ratio of 
 of galaxy to matter power spectra, 
is constant, i.e. scale independent.
On  scales of
$0.02 < k < 0.15 \hompc$ 
the fluctuations are close
to the linear regime, and there are theoretical  reasons 
(e.g. Fry 1996; Benson et al. 2000)
to expect that on large scales 
the biasing parameter 
should tend to a constant and close to unity at the present epoch. 
This is supported by the derived biasing close to unity by combining 
2dFGRS with the CMB (Lahav et al. 2002) and by the 
study of the bi-spectrum of the 2dFGRS alone (Verde et al. 2002).

The 2dFGRS power spectrum (Figure 1) was fitted in Percival et al. (2001) 
over the above range in $k$, 
assuming scale-invariant primordial 
fluctuations and a $\Lambda$-CDM cosmology, for 
four free parameters: $\omegam h$, $\omegab/\omegam$, $h$  
and the redshift space $\sigma^S_{8{\rm g}}$.
The amplitudes
of the linear-theory rms fluctuations are traditionally labeled  $\sigma_{8{\rm m}}$ 
in mass  $\sigma_{8{\rm g}}$ in galaxies, defined on $8 \hompc$ spheres.
Assuming a Gaussian prior on the Hubble constant $h=0.7\pm0.07$ (based
on Freedman et al. 2001) the shape of the recovered spectrum within
the above $k$-range was used to yield 68 per cent confidence limits on
the shape parameter $\omegam h=0.20 \pm 0.03$, and the baryon fraction
$\omegab/\omegam=0.15 \pm 0.07$, in accordance with the popular
`concordance' model (e.g. Bahcall et al. 1999).  For fixed
`concordance model' parameters $n=1, \omegam = 1 - \omegal = 0.3$,
$\Omega_{\rm b} h^2 = 0.02$ and a Hubble constant $h=0.70$, 
the amplitude of 2dFGRS galaxies in redshift space is $\sigma_{8{\rm
g}}^S (L_s,z_s) \approx 0.94$ (at the survey's effective luminosity and redshift).

Recently the SDSS team presented their results for the power spectrum
(Tegmark et al. 2003a,b; Pope et al. 2004), and they found good agreement
with the 2dFGRS gross shape of the power spectrum.
Pope et al. (2004) emphasize that SDSS alone cannot 
break the degeneracy between $\omegam h$ and $\omegab/\omegam$
because the baryon oscillations are not resolved given
window function of the survey.

\section {Upper limit on the neutrino mass} 

Solar, atmospheric, and reactor neutrino experiments have confirmed
neutrino oscillations, implying that neutrinos have non-zero mass, but
without pinning down their absolute masses.  While it is established
that the effect of neutrinos on the evolution of cosmic structure is
small, the upper limits derived from large-scale structure could help
significantly to constrain the absolute scale of the neutrino masses.
Elgar\o y et al. (2002) used the 2dFGRS power spectrum (Figure 1) to
provide an upper limit $m_{\nu,\rm tot} < 2.2\;{\rm eV}$ ,
i.e. approximately 0.7 eV for each of the three neutrino flavours, or
phrased in terms of their contribution to the matter density,
$\Omega_{\nu} / \Omega_{\rm m} < 0.16$.

The WMAP team (Spergel et al. 2003)  reported an improved
limit of $m_{\nu,\rm tot} < 0.71\;{\rm eV}$ (95\% CL).  However, we point out
that neutrinos with eV masses are basically indistinguishable from
cold dark matter at the epoch of last scattering, and therefore they
have little effect on the CMB fluctuations.  The main neutrino signature comes from
the 2dFGRS and the Lyman $\alpha$ forest which were combined with the
WMAP data. The contribution of WMAP is that it constrains better the
other parameters involved, e.g. $\omegam$ (see also Hannestad 2003 and
Tegmark et al. 2003b for similar results from SDSS+WMAP).
Despite the uncertainties involved, it is remarkable that the results from
redshift surveys give upper limits which are lower than 
those deduced from laboratory experiments, e.g. tritium decay.

As the suppression of the power spectrum depends on the ratio
$\Omega_{\nu}/\omegam$, Elgar\o y \& Lahav (2003) found that the
out-of-fashion Mixed Dark Matter (MDM) model, with $\Omega_{\nu}=0.2$,
$\omegam=1 $ and no cosmological constant, fits the 2dFGRS power
spectrum well, but only for a Hubble constant $H_0 < 50\;{\rm km}\,{\rm
s}^{-1}\,{\rm Mpc}^{-1}$.  Blanchard et al. (2003) reached a similar
conclusion, and they also found that the CMB power spectrum could be
fit well by the same MDM model if one allows features in the
primordial power spectrum.  It is intriguing (and perhaps
disappointing) that the CMB and redshift surveys cannot on their own
(i.e. without a strong prior on the Hubble constant) `prove' the
existence a non-zero Cosmological Constant.  Another consequence of
this is that excluding low values of the Hubble constant, e.g. with
the HST Key Project, is important in order to get a strong upper limit
on the neutrino masses.

\begin{figure}
\epsfig{figure=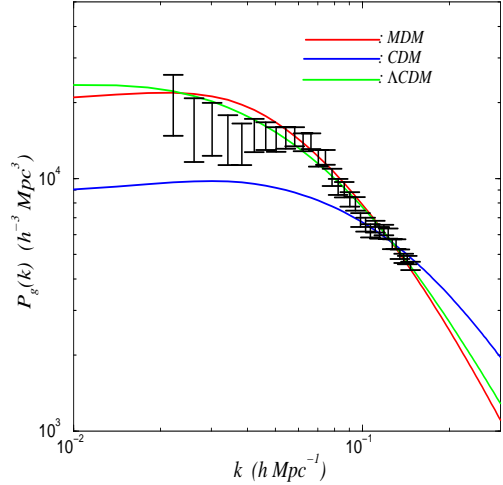,height=3truein,width=3truein}
\caption{The observed 
2dFGRS power spectrum (in redshift space and convolved  with the 
survey window function; Percival et al. 2001) contrasted 
with models.
The three models are 
the old Cold Dark Matter  model
($\Omega_{\rm m}=1$, $\Omega_\nu=0$, $h=0.45$, $n=0.95$),
the `concordance' model  
$\Lambda {\rm CDM}$ ($\Omega_{\rm m}=1$, $\Omega_\Lambda=0.7$, 
$\Omega_\nu = 0$, $h=0.7$, $n=1.0$) 
and  Mixed Dark Matter
 ($\Omega_{\rm m}=1$, $\Omega_\nu=0.2$, $h=0.45$, $n=0.95$),
all with $\Omega_{\rm b}h^2 
= 0.024$.
The models were normalized  to each 
data set separately, but otherwise these are assumed models, not 
formal best fits.
Only the range $ 0.02 < k < 0.15 \hompc$  is used 
at the present linear theory analysis.
These scales of $k$ 
roughly correspond to CMB harmonics $200 <  \ell < 1500$
in a flat $\omegam = 0.3$ universe.
From Elgar\o y \& Lahav (2003).}
\label{pk_mdm}
\end{figure}

\section {The bimodality of galaxy populations}

Madgwick et al. (2002) have utilized the method of Principal 
Component Analysis (PCA) to compress each galaxy spectrum
into one quantity, $\eta \approx 0.5\;pc_1 + pc_2$.
It turns out that  $\eta$ is a useful indicator of the star formation rate
in a galaxy (Madgwick et al. 2003a).
This allows us to divide the 2dFGRS into $\eta$-types, 
and to study e.g. luminosity functions  
and clustering per type.
Figures 2 and 3 show the bimodality 
in this spectral parameter and in the colour distribution (Peacock 2003),
respectively.
Bimodality is also seen clearly in the SDSS photometric and spectroscopic galaxy data
(Blanton et al. 2003, Kauffmann et al. 2004).
While the concept of two major galaxy populations in the Universe
was recognized long time 
ago by Hubble and others, 2dFGRS and SDSS provide quantitative measures 
of the frequency distribution using objective physical measures
like spectral features and colours.
The details of these distribution functions pose challenges to 
models of galaxy formation, in particular regarding the role
of feedback mechanisms and  the `nature' versus `nurture' question.


\begin{figure}
\psfig{figure=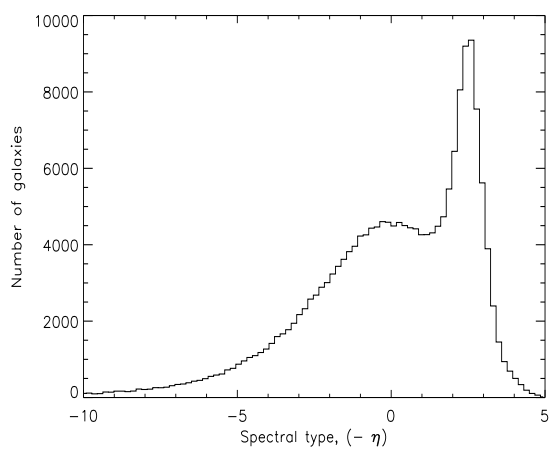,height=2.5truein,width=3truein}
\caption{
  The distribution of PCA spectral type  for 
  2dFGRS galaxies. The distinction between passive `early type' galaxies 
  (right) 
  and actively
  `late type' star-forming galaxies (left) is clear, with a 
  a `valley' centred at 
  $\eta = -1.4$. From Wild et al. (2004).}
\label{fig_eta}
\end{figure}

\begin{figure}
\psfig{figure=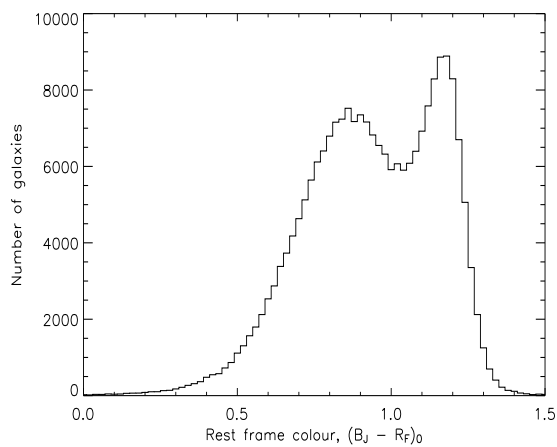,height=2.5truein,width=3truein}
\caption{
  The distribution of rest frame colour for 
  2dFGRS galaxies. The distinction between passive `early type' galaxies 
  (right) 
  and actively `late type' galaxies star-forming galaxies (left) 
  is similar to Figure 2, with 
  a `valley' centred at 
   $(B-R)_0 = 1.07$. From Wild et al. (2004).}
\label{fig_col}
\end{figure}

\section {Clustering per spectral type} 

Although galaxy biasing was commonly neglected until the early 1980s,
it has become evident that on scales $ \simlt 10 \Mpc$  
different galaxy populations exhibit 
different clustering amplitudes, the so-called
morphology-density relation (e.g. Dresser 1980; Hermit et al. 1996). 
Biasing on small scales is also predicted in the simulations
of hierarchical clustering from CDM initial conditions 
(e.g. Benson et al. 2000).
It is important therefore to pay attention to the scale 
on which biasing operates.

Norberg et al. (2002) found that for $L_*$ galaxies, the real space
correlation function amplitude of $\eta$ early-type galaxies is $\sim
50 \% $ higher than that of late-type galaxies.  Peacock et al. (2001), Hawkins et al. (2003)
and Madgwick et al. (2003b) analysed the redshift space correlation
function $\xi(\sigma,\pi)$
in terms of the line-of-sight and perpendicular to the
line-of-sight separation 
for the entire galaxy
populations, we well as for the most passively (`red') and actively
(`blue') star-forming galaxies separately.  The clustering properties
of the two samples are quite distinct on scales $\simlt 10 \Mpc$.  The
`red' galaxies display a prominent `finger-of-god' effect and also
have a higher overall normalization than the `blue' galaxies.  
Figure 4 shows the real
space correlation functions for the red and blue galaxies. While both
are power laws, the slope is different, in accord with results for
populations divided by colour in the SDSS (Zehavi et al. 2002).
Understanding the difference in slope is another challenge for galaxy
formation models.

Biasing could be  non-linear and `stochastic', in the sense that  the number 
of galaxies predicted in a volume is not only a function of the mass
fluctuation in that cell, but is possibly affected by other `hidden variables' 
(Dekel \& Lahav 1999). 
Wild et al. (2004) found recently evidence for a small amount 
stochasticity when considered a joint counts in cells of 
two galaxy populations defined by either colour or spectral type.
The small amount of observed stochasticity supports the use of redshift surveys
for measuring matter density fluctuations on large scales.
However, the comparison with theory calls for  
better understanding of the `hidden variables' in models of galaxy formation
(e.g. Blanton et al. 2000; Somerville et al. 2001).

\begin{figure}
\protect\centerline{
\psfig{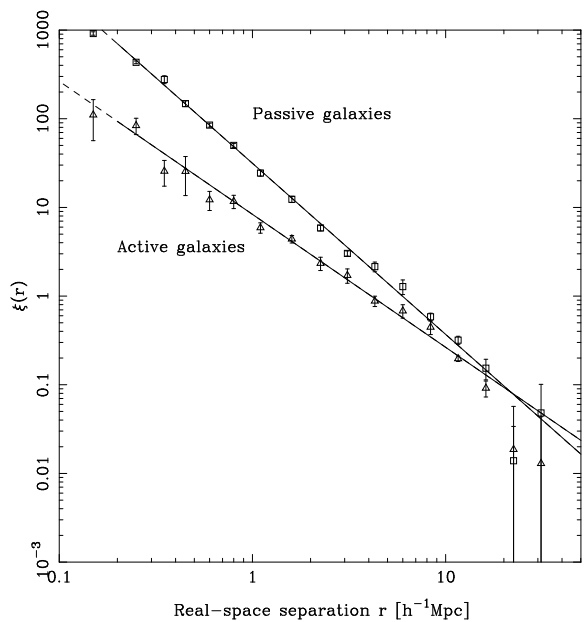}}
 \caption{The non-parametric estimates of the real-space correlation
functions are shown for both our spectral types).  
The solid lines are the best-fitting power
law fits, whereas the dashed lines are
extrapolations of these fits. From Madgwick et al. (2003b).} 
\label{xieta}
\end{figure}

\section {Discussion} 

The results presented  above illustrate the power of redshift surveys 
to address fundamental issues in galaxy formation and Cosmology.
These are only a few examples of the results from the 2dFGRS.
Other results and papers are listed on the 2dFGRS website.

Overall, the results from 2dFGRS fit well
into the `concordance' model which has emerged 
from various cosmological data sets.
The $\Lambda$-CDM model
with comparable amounts of dark matter and dark energy is rather esoteric,
but it is remarkable that different measurements
converge to the `concordance model' with parameters.
Perhaps the least accurate estimates on that list are for $\omegam$ 
and $\sigma_{8{\rm m}}$ (e.g. Bridle et al. 2003, Lahav \& Liddle 2003).
It is intriguing that an Einstein-de Sitter 
 Mixed Dark Matter model (Cold+Hot dark matter) without 
a cosmological constant  can also fit the data, but it requires 
a low Hubble constant and admittedly is at odds with the SN Ia,
cluster baryon fraction  
and other cosmic measurements. It is however an illustration that other
yet unknown models may fit the data equally well.

It may well be that in the future the
cosmological parameters
will be fixed by the CMB, SN Ia etc.
Then, for fixed cosmological parameters,
one can use redshift surveys primarily 
to study galaxy biasing and evolution with cosmic epoch.

\bigskip
\bigskip

\section{Acknowledgements}

I'd like to thank Joss Bland-Hawthorn
and the `Tully 60' conference organisers for the hospitality in Sydney.  
Many of us are grateful to Brent Tully for his numerous original and inspiring 
contributions to the exploration of the local universe.

I thank
Sarah Bridle, \O ystein Elgar\o y, Pirin Erdogdu, Darren Madgwick,
Vivienne Wild and members of the 2dFGRS and Leverhulme Quantitative
Cosmology teams for their contribution to the work summarized here.
The 2dF Galaxy Redshift Survey was made possible through the dedicated
efforts of the staff of the Anglo-Australian Observatory.  I also
acknowledge a PPARC Senior Research Fellowship.

 \label{lastpage}

\end{document}